\documentstyle[psfig,12pt]{article}
\newcommand{\eut}{
\begin{picture}(6,3)(-2,-2)
\put(1,-15){$\tilde{}$}
\put(-2,-2){$\eta$}
\end{picture}}

\begin{document}
\thispagestyle{empty}

\centerline {\bf MINISUPERSPACE EXAMPLES OF QUANTIZATION USING}

\centerline {\bf CANONICAL VARIABLES OF THE ASHTEKAR TYPE:}

\centerline {\bf STRUCTURE AND SOLUTIONS}
\vskip 1 true cm

\centerline{J. Fernando Barbero$^*$}
\centerline{and}

\centerline{Michael P. Ryan, Jr.$^{\dagger}$}
\vskip 1 true cm

\centerline {\it Center for Gravitational Physics and Geometry}

\centerline {\it Physics Department}

\centerline {\it Pennsylvania State University}

\centerline {\it University Park PA 16802}
\vskip 1 true cm

\noindent ABSTRACT: The Ashtekar variables have been use to find a
number of exact solutions in quantum gravity and quantum cosmology.
We investigate the origin of these solutions in the context of a
number of canonical transformations (both complex and real) of the
basic Hamiltonian variables of general relativity.  We are able to
present several new solutions in the minisuperspace (quantum
cosmology) sector.  The meaning of these solutions is then discussed.
\vfill

$^*$ Permanent address: Laboratorio de Astrof{\'\i}sica Espacial y
F{\'\i}sica Fundamental\hfil\break (LAEFF), INTA, P.O. Box 50727,
E-28080 Madrid, SPAIN
\vskip 0.25 true cm
$^{\dagger}$ Permanent address: Instituto de Ciencias Nucleares,
UNAM, A. Postal 70-543, M\'exico 04510 D. F., MEXICO
\vfil\eject
\setcounter{page}{1}

\section{Introduction}

\noindent{\it a) General considerations}:

The formulation of the Ashtekar variables\cite{Ash} has led to a considerable
body
of work applying them to problems of quantum gravity.  These
variables are the result of a complex canonical transformation on a set of
3+1 Hamiltonian variables of general relativity related to the ADM
variables.  One point about canonical transformations that is perhaps
slighted is that they sometimes allow one to find particular quantum
solutions.  The usual mapping of quantum solutions from one set of
canonical variables to another is generated by multiplying a solution in
one set of variables, $\Psi$, by $e^{iG}$, where $G$ is the generator of
the particular canonical transformation to give $\psi = e^{iG} \Psi$.
Notice that if
one manages by some technique to find a particular solution $\Psi$ in
one
set of variables and the generating function is known, it is possible to
find a more complicated solution $\psi$ in terms of the old variables.
This technique has been used with the Ashtekar variables to find a few
exact solutions in quantum cosmology \cite{Kod}\cite{MRY}\cite{AP}.

There are a number of ways in which this concept can be extended beyond
these results.  One is to attempt to promote quantum cosmology solutions
to full quantum gravity solutions as was done with the Chern-Simons
solution in terms of the Ashtekar variables\cite{Kod}.
Another route is to study
the concept of canonical transformations on the usual 3+1 variables and
attempt to use the new variables to generate new exact solutions to
quantum gravity.

The plan of this article is to investigate several canonical
transformations related to the transformation that gives the Ashtekar
variables and use them is the sense mentioned above to generate new exact
solutions.  At this point all the solutions we have been able to find are
minisuperspace (quantum cosmology) solutions, specifically solutions for
diagonal Bianchi type IX cosmological models (Mixmaster models). We will
begin with a brief discussion of canonical transformations in Hamiltonian
formulations of general relativity and their relation to quantum
gravity.  We will then use these concepts to write down a series of
equations that can be solved to give families of exact solutions.  Next
we will discuss the form and meaning of these solutions.  The last
section of the article will be devoted to the significance of our
solutions in the context of quantum gravity and suggestions for using
canonical transformations to find new solutions in quantum gravity.
\vskip 20 pt

{\it b) Canonical transformations and Ashtekar Variables}:

Both the ADM Hamiltonian variables for the gravitational field and the
Ashtekar variables are based on a 3+1 decomposition of the four-metric
of space time.  The ADM variables consist of the three-metric components
$g_{ab}$ on $t$ = constant slices, and their conjugate momenta $\pi^{ab}$
constructed from the extrinsic curvature of these surfaces, $K^{ab}$.  The
Ashtekar variables result from a ``Bargmannization''\cite{Bargsy}
of these variables that
is based on a complex canonical transformation similar to $p \rightarrow
f(q) + ip$, $q \rightarrow q$ in an ordinary one-dimensional classical
system.  It is useful to look at a similar transformation in terms of the
ADM variables.  If we write the ADM action as
$$I = \int \{\pi^{ab} \dot g_{ab} - N\left({{1}\over {\sqrt{g}}}G_{abcd}
\pi^{ab}\pi^{cd} - \sqrt{g}\;(\zeta {\;^3\!R}-2\Lambda)\right ) +
N_a \pi^{ab}_{\;\;\;\; |b} \}
d^4 x, \eqno (1.1)$$
where, as usual, ${^3R}$ is the Ricci scalar on $t$ = const. surfaces, $|$
is a covariant derivative on these surfaces, $\Lambda$ is the cosmological
constant, and $G_{abcd}$ is the DeWitt
metric, $G_{abcd} = g_{ac} g_{bd} - (1/2)g_{ab} g_{cd}$. The parameter $\zeta$
is used
to control the signature of space-time; it is chosen to be +1 for Lorentzian
signatures and -1 for
Euclidean signatures. It is possible to
attempt the complex canonical transformation
$$g_{ab} \rightarrow g_{ab}, \eqno (1.2a)$$
$$\pi^{ab} \rightarrow \tau^{ab}, \qquad \tau^{ab} = f^{ab}(g_{cd}) +
i\pi^{ab}. \eqno (1.2b)$$
Here we assume that $f^{ab}$ could depend explicitly on $g_{ab,c}$ as
well as $g_{ab}$. The transformation  is canonical (ignoring topological
complications
in the space of metrics) if and only if
$$\frac{\delta f^{ab}}{\delta g_{cd}}-\frac{\delta f^{cd}}{\delta
g_{ab}}=0. \eqno
(1.3) $$
The action now becomes
$$I = \int \{ \left ({{1}\over {i}} \right ) \tau^{ab} \dot g_{ab} +
i(f^{ab} \dot g_{ab} ) - N\big ( {{1}\over {\sqrt{g}}} [-G_{abcd} \tau
^{ab} \tau^{cd} + 2G_{abcd} f^{ab} \tau^{cd}  - $$
$$- G_{abcd} f^{ab} f^{cd} ]
- \sqrt{g} (\zeta {\;^3\!R}-2\Lambda) \big )
-2iN_a (\tau^{ab}_{\;\;\;\;|b} - f^{ab}_{\;\;\;\;|b}) \} d^4 x.
\eqno (1.4)$$
Notice that the term $if^{ab} \dot g_{ab}$ is a total time derivative because
(1.3)
implies that
$f^{ab}$ is the functional derivative of a functional $S$, i.e. $f^{ab} =
\delta S /\delta g_{ab}$.  It is also possible to remove the three-curvature
term from the Hamiltonian constraint if we take
$$G_{abcd} {{\delta S}\over {\delta g_{ab}}} {{\delta S}\over {\delta g_{ab}}}
 + g \zeta {\;^3\!R} = 0, \eqno (1.5)$$
which is the Einstein-Hamilton-Jacobi equation\cite{MTW} for $S$ with
the sign of $g{\;^3\!R}$ reversed.

There are a number of observations that we can make about the action (1.4).
First of all, if $S$ is taken to satisfy (1.5), the Hamiltonian
constraint,
$${\cal H}_{\perp} = 0 = -G_{abcd}\tau^{ab} \tau^{cd} + 2G_{abcd}
{{\delta S}\over {\delta g_{ab}}}\tau^{cd}+2 g \Lambda, \eqno (1.6)$$
is an algebraic function of $\tau^{ab}$ with at most second-order terms.
An obvious solution to ${\cal H}_{\perp} = 0$
(when $\Lambda=0$) is $\tau^{ab} =
0$,
which is
the complex equivalent of the usual Einstein-Hamilton-Jacobi formulation
where from (1.2b) $\pi^{ab} = i\delta S /\delta g_{ab}$, $S$ obeys
(1.5), and the space constraint reduces to $(\delta S/\delta g_{ab})_{|b}
= 0$.

Notice that the complex version of the canonical transformation is
not necessary, one can define $\pi^{ab}
\rightarrow  \tau^{ab} = \delta S/\delta g_{ab} + \pi^{ab}$, and the action
becomes
$$ I = \int \{ \tau^{ab} \dot g_{ab} - N({{1}\over {\sqrt{g}}} [
G_{abcd} \tau^{ab} \tau^{cd} - 2G_{abcd} {{\delta S}\over {g_{ab}}}
\tau^{cd} +$$
$$ + G_{abcd} {{\delta S}\over {\delta g_{ab}}} {{\delta S}\over
{\delta g_{cd}}} ] - \sqrt{g}(\zeta {\;^3\!R}-2\Lambda)) +
2N_a[\tau^{ab}_{\;\;\;\;|b} + \left (
{{\delta S}\over {\delta g_{ab}}}\right )_{|b} ] \} d^4 x, \eqno (1.7)$$
and if $S$ obeys the ordinary Einstein-Hamiltonian-Jacobi equation
$$G_{abcd} {{\delta S}\over {\delta g_{ab}}} {{\delta S}\over {\delta g_{cd}}}
- g \zeta\;{^3\!R} = 0, \eqno (1.8)$$
$\tau^{ab} = 0$ is a solution to the Hamiltonian constraint (for $\Lambda=0$)
and $(\delta S/\delta g_{ab})_{|b} = 0$ is again the content of the space
constraint.
Of course it is not necessary to assume that $\tau^{ab}$ is equal to zero,
or that $S$ be a solution to the Einstein-Hamilton-Jacobi equation.  If
$S$ is assumed to be any function of $g_{ab}$ (and its derivatives),
then, for example, in (1.7) the Hamiltonian constraint becomes
$${\cal H}_{\perp} = {{1}\over {\sqrt{g}}}[G_{abcd} \tau^{ab} \tau^{cd}
- 2G_{abcd} {{\delta S}\over {\delta g_{ab}}} \tau^{cd}] - \sqrt{g}\zeta\;
({^3\!R}^{\prime}-2\Lambda), \eqno (1.9)$$
where ${^3\!R}^{\prime}$ is a new ``scalar curvature'' defined by
$${^3R}^{\prime} = {^3R} - {{1}\over {\zeta g}} G_{abcd} {{\delta S}\over
{\delta
g_{ab}}} {{\delta S}\over {\delta g_{cd}}}. \eqno (1.10)$$

In view of the fact that there are a number of possible linear combinations
of $\pi^{ab}$ and $\delta S/\delta g_{ab}$ that are, in principle,
acceptable, we would like to study general transformations of the form
$$\pi^{ab} \rightarrow \tau^{ab} = \delta S/\delta g_{ab} + \beta \pi^{ab},
\eqno (1.10a)$$
$$g^{ab} \rightarrow g^{ab}, \eqno (1.10b)$$
where $S$ will not be assumed to be a solution of the resulting
Hamilton-Jacobi equation and $\tau^{ab}$ will
not necessarily be taken to be zero.
Which of the approaches outlined above one chooses depends on the system one
is studying and the goal one is trying to achieve.  For the classical
theory it might seem to be less desirable to use an $S$ that does not
remove the curvature term from the Hamiltonian constraint, but if one is
using a complex canonical transformation, the quantum theory can be made
more difficult by the necessity of imposing a reality condition on quantum
states, and one can trade the existence of ${^3R}^{\prime} \neq 0$ for
explicitly real quantum variables\cite{fer}.

Notice that even for complex canonical transformation we mentioned above
there are, in principle, as many such transformations as there are solutions
$S$ to (1.5).  The main difficulty in finding functions, $S$, of $g_{ab}$ and
$g_{ab,c}$ is that $f^{ab} = \delta S/\delta g_{ab}$ is a two-index
object, and it is difficult to construct such an object that satisfies
(1.5) solely from $g_{ab}$ and $g_{ab,c}$ (or the Christoffel symbols
$\Gamma^a_{bc}$).  However, if one considers an orthonormal basis of
one-forms on $t =$ const. surfaces,
$\sigma^i = e^i_a (x^c) dx^a$, ($ds^2 = \sigma^i \sigma^i$),
then the connection coefficients $\Gamma^i_{jk}$ have the natural symmetry
$\Gamma_{ijk} = \Gamma_{[ij]k}$ and it is possible to construct the spin
coefficients $\Gamma^i_j = (1/2) \varepsilon ^{i\ell k} \Gamma_{\ell kj}$.
Using these, Ashtekar was able to find an elegant solution for the
equivalent of $S$.

Before we continue, we introduce some notation that will be used
throughout the paper. $SO(3)$ indices will be denoted by lower case
latin letters from the middle of the alphabet, $i,\,k,\,\ell\cdots =1,\,2,\,3$
(we reserve
letters from the beginning of the alphabet for tangent space indices). We will
use
additional indices $I, \,J,\,K,\cdots=1,\,2,\,3$ as labels for certain
geometrical
objects. The 3-dimensional Levi-Civita tensor density and its inverse are
denoted as
$\tilde \eta^{abc}$ and $\eut_{abc}$ and the
internal Levi-Civita tensors will be denoted as
$\varepsilon_{IJK}$, $\varepsilon_{ijk}$. The basic fields in the ADM formalism
with an
$SO(3)$ internal symmetry are a densitized triad ${\tilde E}^{a}_{i}$ and an
object
$K_{a}^{i}$ closely related to the extrinsic curvature. We introduce
$e_{i}^{a}$
as the inverse of $e_{a}^{i}$ (the coefficients of $\sigma^i$) satisfying
$e^a_i e^i_b = \delta^a_b$, $e^a_i e_a^j = \delta_i^j$. The
determinant of $e^i_a$ is defined as,
$${\rm det} e^i_a \equiv \tilde e = {{1}\over {6}} \tilde \eta^{abc}
\varepsilon_ {ijk} e^i_a e^j_b e^k_c. \eqno (1.11)$$
Finally, the SO(3) connection $\Gamma^i_a$ (where $\Gamma^i_a = \Gamma^i_j
e^j_a$,
$\Gamma^i_j$ defined above) compatible with $e^i_a$ is
$$\Gamma^i_a = -{{1}\over {2\tilde e}} (e^i_a e^j_b - 2 e^j_a e^i_b)
\tilde \eta^{bcd} \partial_c e_{dj}. \eqno (1.12)$$
The Ashtekar variables are a densitized basis $\tilde E^a_i \equiv
2\tilde e e^a_i$ and the equivalent of $\tau^{ab}$, $A^i_a = \Gamma^i_a +
iK^i_a$.  We will not go into the details
of of the ADM action written in terms of these variables (although we will give
the
constraints later) but just point out that $\tilde E^a_i$
and $A^a_i$ are new canonical variables derived from $\tilde E^a_i$ and $K^i_a$
by means of a complex canonical transformation generated by the
equivalent of $S$, $\tilde S = 2i\int \tilde E^a_i \Gamma^i_a d^3 x$.
Notice that we have an explicit form for $\tilde S$, whereas in the
previous formulation embodied in (1.2) we would have had to find a
solution
to the Einstein-Hamilton-Jacobi equation in order to give an explicit
expression
for $S$.  Introducing $\tilde E^a_i$ as a basic variable introduces a new
symmetry,
the freedom to perform SO(3) rotations in the $ijk$-indices without
changing $g_{ab}$.  This symmetry is mirrored in a new constraint which
needs to be added to the usual diffeomorphism and Hamiltonian constraints.
The constraint structure in the new variables is
$$\nabla_a \tilde E^a_i = 0, \eqno (1.13a)$$
$$F_{ab}^i \tilde E^a_i = 0, \eqno (1.13b)$$
$$\varepsilon^{ijk} \tilde E^a_i \tilde E^b_j F_{abk} + 2({\rm det} \tilde
E^a_i )\Lambda = 0, \eqno (1.13c)$$
where (1.13a) is the basis-rotation or Gauss-law constraint, (1.13b) is
the diffeomorphism constraint, and (1.13c) is the Hamiltonian constraint
(modulo a term proportional to the Gauss law constraint). The quantity
$$F^i_{ab} = 2\partial_{[a} a^i_{b]} + \varepsilon^i_{jk} A^j_a A^k_b,
\eqno (1.14)$$ is the
curvature of the connection $A^i_a$,
and $\nabla_a \lambda_i = \partial_a \lambda_i + \varepsilon^i_{jk}A_{aj}
\lambda_k$ is the $SO(3)$ covariant derivative acting on internal indices.

Of course, the definition $A^i_a = \Gamma^i_a + iK^i_a$ is equivalent to
$\tau^{ab} = \delta S/\delta g_{ab} - i\pi^{ab}$, and as we mentioned
above,
the possibility of other linear combinations of momenta and functions
of the metric (here triad) still hold.  It should be possible to find a
generator for a real canonical transformation of the form $A^i_a =
-f^i_a + K^i_a$ that removes the potential term from the Hamlitonian
constraint just as the transformation $\tau^{ab} = -\delta S/\delta g_{ab}
+ \pi^{ab}$ does if the generator $S$ satisfies the equivalent of the
normal Einstein-Hamilton-Jacobi equation.  Unfortunately, no such generator
is known at this time.  It is also possible to construct general
transformations of the type given by (1.10) which do not remove the
curvature
terms from (1.3).  In fact, one of us (F. B.) has studied canonical
transformations of the type $\tilde E^a_i = \tilde E^a_i$, $A^i_a =
\Gamma^i_a + \beta K^i_a$ where $\beta$ was taken to be -1 and the
Hamiltonian constraint became\cite{fer}
$$\varepsilon^{ijk} \tilde E^a_i \tilde E^b_j (F_{abk} - 2R_{abk})
- ({\rm det} \tilde E^a_i)\Lambda = 0, \eqno (1.15)$$
where $R_{ab}^i$ is the curvature of the connection $\Gamma^i_a$,
$$R_{ab}^i = 2\partial_{[a} \Gamma^i_{b]} + \varepsilon^i_{jk} \Gamma_a^j
\Gamma^k_b. \eqno (1.16)$$
As mentioned above, real transformations have the advantage (at the possible
cost
of more complicated equations) of not requiring reality conditions in
the quantum formulation.
\vskip 1 true cm

\noindent {\it c) Minisuperspace Models\/}
\vskip 0.5 true cm

In order to give concrete examples of possible Hamiltonian formulations
that we discussed above, we would like to apply them to minisuperspace models
where, as we will show, they lead to a number of exact solutions in the
minisuperspace sector, some of which were known and some of which are
new.

The minisuperspace examples we will use are the diagonal Class A Bianchi
cosmological models where the metric has the form
$$ds^2 = -dt^2 + g_{IJ} (t) \sigma^I \sigma^J, \eqno (1.17)$$
where $g_{IJ}$ is a diagonal matrix and the $\sigma^I$ are invariant
one-forms that satisfy
$d\sigma^I = (1/2) C^I_{\;\;JK} \sigma^J \wedge \sigma^k$ where the
$C^I_{\;\;JK}$
are structure constants of the form $C^I_{JK} = m^{IL}\varepsilon_{LJK}$,
where $m^{IJ}$ is a matrix of constants \cite{RandS}.  We will be most
interested
in the Bianchi IX case where $m^{IJ} = \delta^{IJ}$.

For Class A Bianchi models we can write the ADM action in terms of
$g_{IJ}(t)$ and the basis components of the ADM momentum as
$$I = \int \{ \pi^{IJ} \dot g_{IJ} - N\left ({{1}\over {\sqrt{g}}} [
G_{IJKL} \pi^{IJ} \pi^{KL}] - \sqrt{g} (\zeta{^3R}-2\Lambda) \right ) \} dt
\sigma^1 \wedge
\sigma^2 \wedge \sigma^3, \eqno (1.18)$$
since for all vacuum diagonal Class A models the diffeomorphism constraint is
identically
zero.  Here ${^3R}$ is an algebraic function of $g_{IJ}(t)$ and the
structure
constants $C^I_{\;\;JK}$ (or $m^{IJ})$, and $g = {\rm det} (g_{IJ})$.

It is now possible to make the same canonical transformation $\tau^{IJ}
= f^{IJ} + i\pi^{IJ}$ as in (1.2), but the advantage in the minisuperspace
is that we can realize $f^{IJ}$ as $\partial S/\partial g_{IJ}$, replacing
the functional derivative by a partial derivative.  The
Einstein-Hamilton-Jacobi
equation (1.5) becomes a partial differential equation,
$$G_{IJKL} {{\partial S}\over {\partial g_{IJ}}} {{\partial S}\over
{\partial g_{KL}}} + g\zeta{^3R} = 0. \eqno (1.19)$$
This equation has a number of particular solutions for Bianchi IX models,
some of which have been given elsewhere \cite{MRY}\cite{grah} and several more
which will be
discussed below.  Notice again that this equation contains the entire
classical problem (both Lorentzian and Euclidean), so, in principle, it has a
rich
solution space and consequently a large family of canonical transformations of
the form $\tau^{IJ} = (\partial S/\partial g_{IJ}) + i\pi^{IJ}$.

As before, $\tau^{IJ} = 0$ is a solution to the Hamiltonian constraint
($\Lambda=0$)
$${\cal H}_{\perp} = 0 = -G_{IJKL} \tau^{IJ} \tau^{KL} + 2G_{IJKL}
{{\partial S}\over {\partial g_{IJ}}} \tau^{KL}+2g\Lambda, \eqno (1.20)$$
but this is not the only possible solution, and as before, linear
canonical transformations of the form $\tau^{IJ} = (\partial S/\partial
g_{IJ} ) + \beta \pi^{IJ}$ are possible, where $S$ may or may not be
chosen to annihilate the ${^3R}$ term.  we will discuss these possibilities
in the minsuperspace context below.

While we will refer to the connection between the formulation in terms of
the $S$ solutions given above and the Ashtekar variables and similar
variables, we will generally work in terms of the variables themselves in
order to write the minisuperspace quantum equations.  Writing the metric in
the form of (1.7), the one-forms $\sigma^I$ are $\sigma^I = \sigma^I_a(x^c)
dx^a$,
and the orthonormal one-forms $e^i_a$ are $e^i_I(t) \sigma^I_a$.  The
variables given above become:
$$e^i_a = e^i_I(t)\sigma^I_a (x^c), \eqno (1.21a)$$
$$A^i_a = a^i_I(t) \sigma^I_a (x^c), \eqno (1.21b)$$
$$K^i_a = k^i_I(t) \sigma^I_a (x^c), \eqno (1.21c)$$
$$\tilde E^a_i = E^I_i (t) {\rm det} (\sigma)
\sigma^a_I (x^c), \eqno (1.21d)$$
where all the $x$-dependence is contained in $\sigma^I_a (x^c)$.  The
basic variables for our presentation will be functions of $t$ (and $m^{IJ}$)
only that can be written in the form
$$F_{jk}^i = (a^i_I m^{IL} \varepsilon_{LJK} + \varepsilon^i_{rs} a^r_J
a^s_K) e^J_j e^K_k, \eqno (1.22)$$
$$\Gamma^i_{\ell} = -{{1}\over {2({\rm det} e^i_I)}} [\delta^i_{\ell}
e^j_J e_{jK}m^{KJ} - 2e^i_I e_{\ell K} m^{IK}], \eqno (1.23)$$
and the curvature
$$\varepsilon^{ijk} \tilde E^a_i \tilde E^b_j R_{abk} = -2 ({\rm det}
\sigma^I_a)^2 e^i_I e_{iJ} e_K^j e_{Lj} (m^{LK} m^{IJ} - 2m^{LJ} m^{IK})],
\eqno (1.24)$$
which, since it is a density, retains the $x$-dependent determinant $({\rm
det} \sigma^I_a)$.

Diagonal Class A Bianchi models, as we will see below,
satisfy identically the Gauss-law and diffeomorphism constraints,
so the only constraint that survives and provides the quantum operator we
will need to determine the minisuperspace wave function of our models is the
Hamiltonian
constraint.
\vskip 0.5 true cm

\noindent {\it d) The Quantum Problem\/}

As we mentioned above, we would like to use quantum minisuperspaces as
models of quantum gravity in which it is possible to find exact particular
solutions that can be used to compare quantization in the different sets
of canonical variables discussed above. We will use the Dirac scheme of
quantization where we will apply the Hamiltonian constraint to a state
function $\Psi$ and obtain a form of the Wheeler-DeWitt equation for
the function $\Psi$.  In order to do this we will realize some of the
operators in the quantum system as derivative operators on functions of
the others.  In the ADM formulation, for example, the metric $g_{IJ}$
and its conjugate momentum $\pi^{IJ}$ become operators and one can make
$\Psi$ a function of $g_{IJ}$ and realize $\pi^{IJ}$ as $-i\partial
/\partial g_{IJ}$.  It is also possible to choose the ``momentum
representation'' in which $\Psi$ is a function of $\pi^{IJ}$ and
$g_{IJ}$ is realized as $-i\partial /\partial \pi^{IJ}$.  It is also
possible to have $\Psi (g_{IJ})$ and realize $\tau^{IJ}$ as $\partial
/\partial g_{IJ}$ as in the Bargmann-Segal formulation\cite{Bargsy}
or choose the
``connection representation'' in which $\Psi = \Psi (\tau^{IJ} )$ and
$g_{IJ}$ becomes $-\partial /\partial \tau^{IJ}$.  Since $a^i_I$ in
the Ashtekar representation is the conjugate of $E^I_i$, we have the
same possiblities, that of $E^I_i \rightarrow E^I_i$, $a^i_I \rightarrow
\partial /\partial E^I_i$ or $E^I_i \rightarrow \partial /\partial a^i_I$,
$a^i_I \rightarrow a^i_I$.  We will investigate a number of them.

Notice that for {\it any\/} solution to (1.19) the operator version
of (1.20) in the $\tau_{IJ} \rightarrow \partial /\partial g_{IJ}$
representation (with $\Lambda = 0$ and all derivatives standing to
the right) has the form
$$\hat {\cal H}_{\perp} = -G_{IJKL} {{\partial ^2}\over {\partial
g_{IJ} \partial g_{KL}}} + 2G_{IJKL} {{\partial S}\over {\partial
g_{IJ}}} {{\partial}\over {\partial g_{KL}}}, \eqno (1.25)$$
which has as a solution to $\hat {\cal H}_{\perp} \Psi = 0$, $\Psi
= \Psi_0 =$ const.  By the usual transformation of variables we can
construct a solution of the usual Wheeler-DeWitt equation of the form
$\psi = e^{\pm S} \Psi_0$.  This formal solution becomes a true
solution if we have an explicit solution for $S$.  Notice that for
the usual
Hamilton-Jacobi formulation a solution $\psi = e^{\pm iS} \Psi_0$
is possible if a solution $S$ can be found for the usual Einstein-
Hamilton-Jacobi equation.   It is also possible to study the quantum
problem for any of the linear canonical transformations of the form
$g_{IJ} \rightarrow g_{IJ}$, $\pi^{IJ} \rightarrow f^{IJ} + \beta
\pi^{IJ}$.  Of course, the equivalent Ashtekar-like transformations
also lead to quantum equations and there exist similar maps among the
quantum solutions.  Our plan is to investigate a number of these
possibilties, present several exact solutions, and use them as
model examples of possible quantum solutions in the full theory of
gravity and discuss the relation between them and such solutions.

\section{Solutions to the Hamilton--Jacobi Equation}

The purpose of this section is to discuss new solutions to the Hamilton--Jacobi
equations as an intermediate step to finding solutions to the Wheeler--DeWitt
equation for Bianchi IX models. We do this both in the real ADM and the real
Ashtekar
formulations.
Whereas in the ADM case the theory is explictly real for both Euclidean and
Lorentzian signatures (that we describe  collectively by using the parameter
$\zeta$
introduced above) the usual way to treat Lorentzian signatures with Ashtekar
variables is by working with complex fields and imposing some ``reality"
conditions
that can be used, for example, as a tool for fixing the scalar product. A less
conventional attitude is to use explicitly real ``Ashtekar-like" variables (in
the sense that they
keep their geometrical meaning)\cite{fer}.
The Lorentzian theory is
recovered by modifying the Hamiltonian constraint through the introduction of a
potential term. In this paper we choose to concentrate on this second, and more
novel,
approach, motivated by the desire to know whether this new formulation provides
us with a useful alternative to the use of reality conditions. To this end it
proves to
be convenient to compare the results obtained with ones corresponding to the
ADM
case, so we begin by studying the geometrodynamical formulation.
In order to facilitate the comparison between the ADM and Ashtekar formalisms
we
slightly modify  the ADM constraints by introducing an internal $SO(3)$
symmetry. For
the specific example of the Bianchi IX model we get the constraints
$$\begin{array}{l}
\varepsilon_{ijk}k_{I}^{j}{\tilde E}^{kI}=0\\
\varepsilon^{I}_{\;\;JK}
k_{I}^{j}E_{i}^{J}-2\varepsilon_{ijk}e_{I}^{j}e_{Nk}\delta^{IN}k_{k}^{i}=0\\
(det \sigma)^{2}\left\{2\zeta\left[(Tr e^{2})^{2}-2(Tr
e^{4})\right]-2\Lambda(det
E_{i}^{I})+2k_{I}^{[i}k_{J}^{j]}E^{I}_{i}E_{j}^{J}\right\}=0,
\end{array}
\eqno (2.1)$$
where $Tr e^{2}\equiv e_{I}^{i}e_{J}^{j}\delta^{IJ}\delta_{ij}$ and $Tr
e^{4}\equiv
e_{I}^{i}e_{J}^{j}e_{K}^{k}e_{L}^{l}\delta^{IJ}\delta^{KL}\delta_{jk}\delta_{il}$ and
$\zeta$ controls the space--time signature ($\zeta=\pm 1$ for Lorentzian and
Euclidean signatures respectively). In
this paper we concentrate on the study of Mixmaster models for which
$k_{I}^{i}$ and
${\tilde E}_{i}^{I}$ are taken to be diagonal,
$$\begin{array}{ll}
k_{I}^{i}\equiv\left[
                     \begin{array}{lll}
                         \mu & 0 & 0 \\
                          0 & \nu &0 \\
                          0 & 0 & \lambda
                     \end{array}
               \right],
&
E^{I}_{i}\equiv\left[
                     \begin{array}{lll}
                         M & 0 & 0 \\
                         0 & N & 0 \\
                         0 & 0 & L
                     \end{array}
               \right],
\end{array}
\eqno (2.2)$$
the variables $\mu$, $\nu$, $\lambda$, M, N, L introduced above are canonically
conjugate pairs,
i.e.
$$\begin{array}{l}
\{M, \mu\}=1,\\
\{N, \nu\}=1,\\
\{L, \lambda\}=1,
\end{array} \eqno (2.3)$$
and the remaining Poisson brackets are zero. The previous expressions, together
with
(1.21d) allow us to write
$$
e_{I}^{i}=\frac{1}{\sqrt{2}}(L M N)^{1/2}\left[
                     \begin{array}{lll}
                         1/M & 0 & 0 \\
                          0 & 1/N &0 \\
                          0 & 0 & 1/L
                     \end{array}
               \right].
\eqno (2.4)$$
Introducing (2.2) and (2.4) in (2.1) we find that the first two constraint
equations are identically satisfied and the scalar constraint is given by
$$\begin{array}{l}
2(\mu\nu MN+\mu\lambda ML+\nu\lambda
NL)+\zeta(M^{2}+N^{2}+L^{2})-\hspace{3cm}\\
\hspace{3cm}-\frac{\zeta}{2}\left(\frac{M^{2}L^{2}}{N^{2}}+
\frac{N^{2}L^{2}}{M^{2}}+\frac{M^{2}N^{2}}{L^{2}}\right)-2\Lambda
MNL=0.
\end{array}
\eqno(2.5)$$

The usual Ashtekar constraints for type A Bianchi models are
($m^{IJ}=\delta^{IJ}$ gives
Bianchi IX)
$$\varepsilon_{ijk}
a^{j}_{I}E^{iI}=0, \eqno (2.6a)$$

$$(m^{IL}\varepsilon_{LJK}a^{i}_{I}+
\varepsilon^{ijk}a_{Jj}a_{Kk})E^{K}_{i}=0, \eqno (2.6b)$$

$$\varepsilon_{ijk}\left[a^{i}_{I}m^{IL}
\varepsilon_{LJK}+\varepsilon^{i}_{\;\;\ell
m}a^{j}_{J}a^{k}_{K}\right]E^{J}_{\ell}E^{K}_{m}=0. \eqno(2.6c)$$

With this form of the constraints real variables describe Euclidean signatures
and
complex variables, with the addition of reality conditions, Lorentzian
signatures.
For Bianchi IX diagonal models we write
$$
\begin{array}{ll}
a_{I}^{i}\equiv\left[
                     \begin{array}{lll}
                         \alpha & 0 & 0 \\
                          0 & \beta &0 \\
                          0 & 0 & \gamma
                     \end{array}
               \right],
&
E^{I}_{i}\equiv\left[
                     \begin{array}{lll}
                         A & 0 & 0 \\
                         0 & B & 0 \\
                         0 & 0 & G
                     \end{array}
               \right],
\end{array}
\eqno(2.7)
$$
where the pairs $(\alpha,A)$, $(\beta,B)$, $(\gamma,G)$ are canonically
conjugate
\footnote{For the Euclidean theory this means just that
$\{\alpha,A\}=\{\beta,B\}=\{\gamma,G\}=1$ with the remaining Poisson brackets
equal to
zero. If we consider complex variables the previous Poisson brackets pick a
purely
imaginary factor and become $\{\alpha,A]\}=\{\beta,B\}=\{\gamma,G\}=i$.}.
If we use the real formulation in terms of Ashtekar-like variables introduced
in
Ref. \cite{fer}
and follow the same steps as before we find that the Hamiltonian constraint has
a
potential term. For the Bianchi IX model the constraint is given by
$$\begin{array}{l}
\varepsilon_{ijk}\left[a_{I}^{i} m^{IL}
\varepsilon_{LJK}+\epsilon^{i}_{\;\;\ell m}
a^{\ell}_{J}a^{m}_{K}\right]E^{J}_{j}E^{K}_{k}-2\Lambda(\det E^{i}_{I})\\
+4(e^{i}_{I}e_{iJ})(e^{j}_{K}e_{Lj})\left[m^{LK}m^{IJ}-2m^{LJ}m^{IK}\right]
=0,
\end{array}
\eqno(2.8)$$
with $m^{IJ} = \delta^{IJ}$. If we introduce (2.7), the
Gauss law and vector constraints are identically satisfied (as before) and the
scalar constraint (2.6c) becomes
$$
2(\alpha\beta+\gamma)AB+2(\alpha\gamma+\beta)AG+2(\beta\gamma+
\alpha)BG+2\Lambda\;ABG=0,
\eqno(2.9)$$
while the constraint (2.8) for the real Ashtekar-like variables
becomes
$$\begin{array}{l}
2(\alpha\beta+\gamma)AB+2(\alpha\gamma+\beta)AG+2(\beta\gamma+\alpha)BG+\hspace{3cm}
\\
+2(A^{2}+B^{2}+G^{2})-\left(\frac{A^{2}B^{2}}{G^{2}}+\frac{A^{2}G^{2}}{B^{2}}+
\frac{B^{2}G^{2}}{A^{2}}\right)-2\Lambda\;ABG=0.
\end{array}
\eqno(2.10)$$

In conclusion, we see that to quantize these models we only
need to consider the Hamiltonian
constraints (2.5), (2.9), and (2.10). In all the cases we can use either a
``position" representation or a momentum representation. Since both
(2.5) and
(2.10) are non-polynomial in some of the variables, we multiply them by
appropriate factors in order to avoid the appearance of derivatives in the
denominators in some of the quantizations of the  model. The unpleasant
consequence of
this is that the differential equations that the wave functions will have to
satisfy are of
very high order.

Starting from (2.5), if we quantize by realizing the operators $\hat
\mu$, $\hat \nu$, $\hat \lambda$, $\hat M$, $\hat N$, $\hat L$ as
$$\begin{array}{l}
{\hat \mu}, {\hat \nu}, {\hat
\lambda}\rightarrow\;\mu,\;\nu,\;\lambda,\\
{\hat M}, {\hat N}, {\hat L}\rightarrow\;
i\partial_{\mu},\;i\partial_{\nu},\;i\partial_{\lambda},\
\end{array}
\eqno(2.11)$$
($\hbar=1$) we have $[{\hat \mu},\;{\hat M}]=-i$, $[{\hat \nu},\;{\hat N}]=-i$,
and
$[{\hat \lambda},\;{\hat L}]=-i$. The Wheeler-DeWitt equation now becomes
(choosing the
operator ordering corresponding to writing all the derivatives to the right and
multiplying the constraint by $N^{2}M^{2}L^{2}$)
$$\begin{array}{l}
\left\{2\left[
\mu\nu\partial_{\mu}\partial_{\nu}+\mu\lambda\partial_{\mu}\partial_{\lambda}+
\nu\lambda\partial_{\nu}\partial_{\lambda}\right]\partial_{\mu}^{2}\partial_{\nu}^{2}
\partial_{\lambda}^{2}+\zeta\partial_{\mu}^{2}
\partial_{\nu}^{2}\partial_{\lambda}^{2}(\partial_{\mu}^{2}+\partial_{\nu}^{2}+
\partial_{\lambda}^{2})-\right.\\
\left.-\frac{1}{2}\zeta(\partial_{\mu}^{4}\partial_{\nu}^{4}+\partial_{\mu}^{4}
\partial_{\lambda}^{4}+\partial_{\nu}^{4}\partial_{\lambda}^{4})-2i\Lambda
\partial_{\mu}^{3}\partial_{\nu}^{3}\partial_{\lambda}^{3}\right\}\Psi=0.\
\end{array}
\eqno(2.12)$$
If instead, we quantize using
$$\begin{array}{l}
{\hat \mu}, {\hat \nu}, {\hat \lambda}\rightarrow\;-i\partial_{M},
\;-i\partial_{N},\;-i\partial_{L},\\
{\hat M}, {\hat N}, {\hat L}\rightarrow\;
M,\;N,\;L,\
\end{array}
\eqno(2.13)$$
we get
$$\begin{array}{l}
\left\{-2(MN\partial_{M}\partial_{N}+ML\partial_{M}\partial_{L}+NL\partial_{N}\partial_
{L})+\zeta(M^{2}+N^{2}+L^{2})-\right.\\
\left.-\frac{1}{2}\zeta\left(\frac{M^{2}L^{2}}{N^{2}}+
\frac{L^{2}N^{2}}{M^{2}}+\frac{M^{2}N^{2}}{L^{2}}\right)-
2\Lambda\;MNL\right\}\Psi=0.
\end{array}
\eqno(2.14)
$$
We consider now the Hamiltonian constraint (2.9). If we quantize according to
$$\begin{array}{l}
{\hat \alpha}, {\hat \beta}, {\hat
\gamma}\rightarrow\;\alpha,\;\beta,\;\gamma,\\
{\hat A}, {\hat B}, {\hat G}\rightarrow\;-i\partial_{\alpha},\;
-i\partial_{\beta}-i\partial_{\gamma},\
\end{array}
\eqno(2.15)$$
and write all the derivatives to the left we get the Wheeler-DeWitt equation
(see
Refs. \cite{Kod}\cite{MRY})
$$
\left[-\partial_{\alpha}\partial_{\beta}(\alpha\beta+\gamma)-
\partial_{\alpha}\partial_{\gamma}(\alpha\gamma+\beta)-
\partial_{\beta}\partial_{\gamma}(\beta\gamma+\alpha)
+i\Lambda\partial_{\alpha}\partial_{\beta}\partial_{\gamma}
\right]\Psi=0.
\eqno(2.16)$$
If we quantize using
$$\begin{array}{l}
{\hat \alpha}, {\hat \beta}, {\hat
\gamma}\rightarrow\;-i\partial_{A},\;
-i\partial_{B}-i\partial_{G},\\
{\hat A}, {\hat B}, {\hat G}\rightarrow\;A,\;B,\;G,
\end{array}
\eqno(2.17)$$
we find (See Ref. \cite{AP})
$$
\!\!\!\!\left[AB(-\partial_{A}\partial_{B}+i\partial_{G})+\!AG(-\partial_{A}\partial_{G}+
i\partial_{B})+\!
BG(-\partial_{B}\partial_{G}+i\partial_{A})+\!\Lambda
ABG\right]\Psi=0.
\eqno(2.18)
$$
The equivlent of this equation for supergravity has been considered
in Ref. \cite{OPR}.  Finally
we consider the new Hamiltonian constraint (2.10). Using the
quantizations introduced above we find, respectively,
$$\begin{array}{l}
\left\{-2\left[(\alpha\beta+\gamma)\partial_{\alpha}\partial_{\beta}+
    (\alpha\gamma+\beta)\partial_{\alpha}\partial_{\gamma}+

(\beta\gamma+\alpha)\partial_{\beta}\partial_{\gamma}\right]\partial_{\alpha}^{2}
    \partial_{\beta}^{2}\partial_{\gamma}^{2}-2i\Lambda\partial_{\alpha}^{3}
    \partial_{\beta}^{3}\partial_{\gamma}^{3}+\right.\\
\left. 2\partial_{\alpha}^{2}\partial_{\beta}^{2}\partial_{\gamma}^{2}
    (\partial_{\alpha}^{2}+\partial_{\beta}^{2}+\partial_{\gamma}^{2})-
    (\partial_{\alpha}^{4}\partial_{\beta}^{4}+\partial_{\alpha}^{4}
\partial_{\gamma}^{4}+\partial_{\beta}^{4}\partial_{\gamma}^{4})\right\}\Psi=0,
\end{array}
\eqno(2.19)$$
and
$$\begin{array}{l}
\left\{2\left[AB(-\partial_{A}\partial_{B}+i\partial_{G})+AG(-\partial_{A}\partial_{G}+
i\partial_{B})+BG(-\partial_{B}\partial_{G}+i\partial_{A})\right]+\right.\\
\left.+2(A^{2}+B^{2}+G^{2})-\left(\frac{A^{2}B^{2}}{G^{2}}+
\frac{A^{2}G^{2}}{B^{2}}+\frac{B^{2}G^{2}}{A^{2}}\right)\right\}\Psi=0,
\end{array}
\eqno(2.20)
$$
where in (2.19) we have multiplied by $A^{2}B^{2}G^{2}$ to avoid derivatives in
the
denominators. Equations (2.17) and (2.19) are ninth  order partial
differential equations (PDE's)(eighth order
if we do not include the cosmological constant term); they are quite
complicated and
will have a number of
``spurious" solutions introduced by multiplying by sixth order polynomials.
Their
solutions must be related to those of (2.14) and (2.20) by Fourier transform.
Equations (2.17) and (2.18) were studied by Kodama\cite{Kod}. In this paper we
will
concentrate on the discussion of the solutions to Eq. (2.20) and their
relationship with solutions to the other equations.

In order to study equations (2.14), (2.16), (2.18), and (2.20)
we will write $\Psi=W e^{-S}$ where W and S are functions of (M, N, L),
($\alpha$,
$\beta$, $\gamma$), (A, B, G), and (A, B, G) respectively. In this way we get
the equations shown in Appendix A. We will look for particular solutions having
the
property that $S$ satisfies a Hamilton--Jacobi equation (much in the spirit of
the WBK
approximation scheme) In this way each of the equations in Appendix A
divides
into two;
an equation for $S$ and another for $W$. The Hamilton--Jacobi equations
obtained
(corresponding to (2.14), (2.16), (2.18), and (2.20) respectively) are
$$\begin{array}{l}
-2\left[MN\partial_{M}S\partial_{N}S+LM\partial_{L}S\partial_{M}S+
LN\partial_{L}S\partial_{N}S\right]+\\
+\left[\zeta(M^{2}+N^{2}+L^{2})-\frac{1}{2}\zeta\left(\frac{M^{2}L^{2}}{N^{2}}+
\frac{N^{2}L^{2}}{M^{2}}+\frac{M^{2}N^{2}}{L^{2}}\right)-2\Lambda\;MNL\right]=0,
\end{array}
\eqno(2.21)$$
$$(\alpha\beta+\gamma)\partial_{\alpha}S\partial_{\beta}S+
    (\alpha\gamma+\beta)\partial_{\alpha}S\partial_{\gamma}S+
    (\beta\gamma+\alpha)\partial_{\beta}S\partial_{\gamma}S
+i \Lambda\;\partial_{\alpha}S\partial_{\beta}S\partial_{\gamma}S=0,
\eqno(2.22)$$
$$\begin{array}{l}
AB\partial_{A}S\partial_{B}S+AG\partial_{A}S\partial_{G}S+BG\partial_{B}S\partial_{G}S+
\\
+i(AB\partial_{G}S+AG\partial_{B}S+BG\partial_{A}S)-\Lambda\;ABG=0,
\end{array}
\eqno(2.23)$$
$$\begin{array}{l}
AB\partial_{A}S\partial_{B}S+AG\partial_{A}S\partial_{G}S+BG\partial_{B}S
\partial_{G}S+\\
+i(AB\partial_{G}S+AG\partial_{B}S+BG\partial_{A}S)-(A^{2}+B^{2}+G^{2})+\\
\frac{1}{2}\left(\frac{A^{2}B^{2}}{G^{2}}+\frac{A^{2}G^{2}}{B^{2}}+
\frac{B^{2}G^{2}}{A^{2}}\right)+\Lambda\;ABG=0.
\end{array}
\eqno(2.24)$$
We will now discuss some solutions (among them several new ones) to the
Hamilton-Jacobi
equations (2.21--2.24). Starting from (2.21), and
putting $\Lambda=0$, we will take an ansatz of the form
$$S=a\left(\frac{LM}{N}+\frac{LN}{M}+\frac{MN}{L}\right)+bL+cM+dN+e,
\eqno(2.25)$$
where $a$, $b$, $c$, $d$, $e$ are constants that we have to fix. The
Moncrief-Ryan
solution\cite{MRY} to
(2.21) is contained in this family and corresponds to $b=c=d=e=0$. Notice that
the addition of linear terms is not trivial due to the non-linear character of
the
Hamilton-Jacobi equation. Substituting (2.25) in (2.21) we find that the
equation is satisfied if the constants in (2.25) are solutions of the following
equations
$$\begin{array}{l}
-4 a^{2}+\zeta=0,\\
bc+2ad=0,\\
bd+2ac=0,\\
cd+2ab=0,
\end{array}
\eqno(2.26)$$
and e is arbitrary. For Lorentzian signatures ($\zeta=+1$) the
possible solutions to
(2.26) are
$$\begin{array}{l}
\!\!\!\!\!\!(a,b,c,d)=(\frac{1}{2},0,0,0),\;(\frac{1}{2},-1,-1,-1),\;(\frac{1}{2},1,1,-1),
\;(\frac{1}{2},1,-1,1),\;(\frac{1}{2},-1,1,1),\\
(-\frac{1}{2},0,0,0),\;(-\frac{1}{2},1,1,1),\;(-\frac{1}{2},-1,-1,1),
\;(-\frac{1}{2},-1,1,-1),\;(-\frac{1}{2},1,-1,-1).\hspace{.7cm}
\end{array}
\eqno (2.27)$$
If $\zeta=-1$ (Euclidean signatures) the solutions to (2.26) are given by
$$\begin{array}{l}
(a,b,c,d)=(\frac{i}{2},0,0,0),\;(\frac{i}{2},-i,-i,-i),\;(\frac{i}{2},i,i,-i),
(\frac{i}{2},i,-i,i),\;(\!\frac{i}{2},-i,i,i),\hspace{.9cm}\\
(-\frac{i}{2},0,0,0),\;(-\frac{i}{2},i,i,i),\;
(-\frac{i}{2},-i,-i,i),
\;(-\frac{i}{2},-i,i,-i),(-\frac{i}{2},i,-i,-i).\hspace{1.6cm}
\end{array}
\eqno(2.28)$$
The solutions shown above seem to be in correspondence with the analytical
solutions
known in closed form for the Bianchi IX model\cite{Bel}.
It is possible that there are no more solutions for $S$ that can be
written in analytic form\cite{belg}.
The  difference between the
solutions for Euclidean and Lorentzian formulations  is the appearance of a
global,
purely imaginary, factor. We have not been able to find solutions for non-zero
cosmological constant.
Equation (2.22) has been studied in detail by Kodama\cite{Kod}. Here we give
the known
solutions for completeness (in this case the cosmological constant $\Lambda$
must be
different from zero)
$$
S=\frac{3i}{2\Lambda}\left[\alpha^{2}+\beta^{2}+\gamma^{2}+2\alpha\beta\gamma\right].
\eqno(2.29)$$
We are not aware of any solution for $\Lambda=0$.

Equations (2.23) and (2.24) differ only in the potential term and the sign of
the
term with the cosmological constant. For $\Lambda=0$ we try solutions of the
form
$$
S=a\left(\frac{AB}{G}+\frac{AG}{B}+\frac{BG}{A}\right)+bA+cB+dG+e.
\eqno(2.30)$$
It is easy to verify that (2.30) are solutions to the Hamilton-Jacobi equation
(2.23) provided that the constants satisfy the following conditions (e
arbitrary)
$$
\begin{array}{l}
a^{2}+ia=0,\\
bc+id+2ad=0,\\
bd+ic+2ac=0,\\
cd+ib+2ab=0,
\end{array}
\eqno(2.31)$$
the solutions to the previous equations are
$$\begin{array}{l}
(a,b,c,d)=(0,0,0,0),\;(0,-i,-i,-i),\;(0,i,i,-i),
(0,i,-i,i),\;(0,-i,i,i),\hspace{.9cm}\\
(-i,0,0,0),\;(-i,i,i,i),\;
(-i,-i,-i,i),\;(-i,-i,i,-i),(-i,i,-i,-i).\hspace{1.6cm}
\end{array}
\eqno(2.32)$$
If we consider instead equation (2.24) with its potential term, (2.30) is a
solution (for $\Lambda=0$) when the constants a, b, c, d, e are solutions to
the
equations
$$
\begin{array}{l}
2a^{2}+2ia-1=0,\\
bc+id+2ad=0,\\
bd+ic+2ac=0,\\
cd+ib+2ab=0,
\end{array}
\eqno(2.33)$$
where, as before, $e$ is arbitrary. The solutions to (2.33) are
$$\begin{array}{l}
\hspace{-1cm}(a,b,c,d)=(\frac{1-i}{2},0,0,0),\;(\frac{1-i}{2},-1,1,1),
\;(\frac{1-i}{2},1,-1,1),
(\frac{1-i}{2},1,1,-1),\\
\hspace{-1cm}\;(\frac{1-i}{2},-1,-1,-1),\;(\frac{i-1}{2},0,0,0),
\;(\frac{i-1}{2},1,1,1),\;
(\frac{i-1}{2},1,-1,-1),\\
\hspace{-1cm}\;(\frac{i-1}{2},-1,1,-1),(\frac{i-1}{2},-1,-1,1).
\end{array}
\eqno(2.34)$$
As before we know of no solution for the $\Lambda\neq0$ case.

\section{Discussion of Solutions}

In order to display the solutions given in the previous section we will
give them in terms of the Misner parametrization of the Mixmaster model
\cite{MTW}.  For the Lorentzian ADM equations (2.1) the metric variables given
in
Eq. (2.2), $M$, $N$, $L$, are, in terms of the Misner variables,
$$g_{11} = e^{2\alpha} e^{2\beta_{+} + 3\sqrt{3} \beta_{-}}, \qquad g_{22} =
e^{2\alpha} e^{2\beta_{+} - 2 \sqrt{3}\beta_{-}}, \qquad  g_{33} =
e^{2\alpha} e^{-4\beta_{+}},$$
$$M = e^{2\alpha}e^{2\beta_{+}},\qquad  N = e^{2\alpha}e^{-\beta_{+} +
\sqrt{3}
\beta_{-}},\qquad  L = e^{2\alpha} e^{-\beta_{+} - \sqrt{3}\beta_{-}}. \eqno
(3.1)$$
This means that the Lorentzian solutions for $S$ given in Eq. (2.25) have
the form (with the trivial constant $e = 0$)
$$S = e^{2\alpha}(a[e^{-4\beta_{+}} + 2e^{2\beta_{+}} \cosh
2\sqrt{3}\beta_{-}] + ce^{2\beta_{+}} +$$
$$+ e^{-\beta_{+}}\{ (b + d)\cosh \sqrt{3} \beta_{-} + (d - b)\sinh
\sqrt{3} \beta_{-}\} ). \eqno (3.2)$$
Notice that the solutions (2.30) for the real Ashtekar equations have
exactly the same form as (3.2) since $A = M$, $B = N$, $G = L$, but with
different constants $a, \cdots, d$.

We would like to display the solutions for $\alpha =$ const. (average
radius of the universe constant) in the $\beta_{+} \beta_{-}$-plane.  The
solution for $S$ (3.2) with $b = c = d = 0$ is, as was mentioned above,
the solution given by Moncrief and Ryan\cite{MRY} (and by Graham as the bosonic
sector of supergravity\cite{grah}) and Figure 1 is a three-dimensional plot of
$|\Psi|^2 = e^{-2S}$ for $\alpha = 0$ in the $\beta_{+} \beta_{-}$-plane.
The line shown on the plot represents a contour of the potential
$V(\beta_{\pm})$ given in Ref. \cite{MTW} which drives the Mixmaster model.
The
potential has a triangular symmetry, where rotation by $\pi/3$ in the
$\beta_{+} \beta_{-}$-plane leaves $V$ invariant.  There are also soft
``channels'' where $V$ goes to zero for large values of $\beta_{+}
\beta_{-}$
that begin at the corners of the triangle and run directly to infinity.
These channels become exponentially narrower at large values of
$\beta_{+}$ and $\beta_{-}$.  The straight lines that define the center
of these channels each represent the Taub model, a special case of the
Mixmaster model\cite{Taub}.  The point $\beta_{+} =
\beta_{-} = 0$ is the $k = +1$ Robertson-Walker universe.  The solution
shown in Fig. 1 is peaked over $\beta_{+} = \beta_{-} = 0$ and has a
roughly triangular form with the points of the triangle lying in the
directions of the three ``Taub'' channels of the potential.

At first glance there seems to be a large number of new solutions with
$b, \cdots, d$ nonzero, but these solutions share the triangular symmetry
of the Misner potential, so some of them are just copies of the others
related by a rotation by $\pi/3$ in the $\beta_{+} \beta_{-}$-plane.  Of
the Lorentzian ``wormhole''\cite{oby} solutions of Eqs. (2.26), with $a = 1/2$,
there is one isolated new solution with $b = c = d = -1$, which has the
triangular symmetry of the potential.  This solution, for $\alpha = 0$ is
shown as a graph of $e^{-2S}$ in Fig. 2.  It has a peak over $\beta_{+} =
\beta_{-} = 0$ as before, but it seems to single out the Taub model
channels with ``arms'' where the solution goes asymptotically to one as
the distance out along the three channels becomes infinite.  The ``arms''
become narrower rapidly as the distance from $\beta_{+} = \beta_{-} = 0$
becomes large.  For the arm along the $\beta_{+}$-axis at $\alpha = 0$
and for small $\beta_{-}$ and large $\beta_{+}$, $e^{-2S}$ becomes
$$\exp\{-12e^{2\beta_{+}} \beta_{-}^2\}, \eqno (3.3)$$
a Gaussian in $\beta_{-}$ with width $\sqrt{1/12}e^{-\beta_{+}}$.  We
will discuss the problem of ``normalization'' of these solutions below.

Of the last three Lorentzian $a = 1/2$ solutions given in (2.17), only
one is relevant, with the others given by $\pi/3$ rotations in the
$\beta_{+} \beta_{-}$-plane.  We choose $b = d = 1$, $c = -1$, and for
these constants $e^{-2S}$ is shown in Figure 3.  This solution is very
strongly peaked over almost the entire Taub model line that is
represented by
$\beta_{-} = 0$.  As before, for large $\beta_{+}$ and small
$\beta_{-}$ we have have exactly the same form for $e^{-S}$ as given in
(3.3) and $e^{-2S} \rightarrow 1$ for $\beta_{-} = 0$ and $\beta_{+}
\rightarrow \infty$.  This solution falls rapidly to zero for $\beta_{+}
< 0$, so the peak near $\beta_{+} = \beta_{-} = 0$
evident in the solutions given
above has disappeared.

For the solution given by (2.30), as we have mentioned, Eq. (3.2) still
describes the ``real Ashtekar'' solutions with $a, \cdots, d$ given by
(2.34).  Since the only difference between this solution and the
Lorentzian solutions is that $a$ becomes complex, but with real part
the same as for the Lorentzian solutions, so $|\Psi|^2$ is the same as in
the Lorentzian case, and Figures 1-3 give this function also.

For the ADM Euclidean solution and the true Ashtekar solutions $S$
becomes pure imaginary, and $|\Psi|^2$ is one, so we give no graphs of
these functions.  Of course, there is no reason to suppose that
$|\Psi|^2$ in the Ashtekar case has any intrinsic meaning (such as the
``probablility'' of finding the universe at some $\beta_{+}\beta_{-}$
point at $\alpha = 0$), since there is no agreement on probability
measures for these complexified theories.

It is obvious, however, that our graphing of $|\Psi|^2$ implies that we
are thinking of the Hartle--Hawking\cite{Hawk} definition of the probablility
associated
with the wave function of the universe.  The definition of probability
measures on solutions to various formulations of quantum gravity has been
a difficult problem.  The Hartle--Hawking definition for ordinary ADM variables
is one possibility, but other definitions in terms of superspace currents
have also been proposed\cite{KR}.  For complexified variables of the
Bargmann-Segal or Ashtekar type, the construction of probability measures
is even more complicated, so we will not attempt to give any such measures
for the solutions given in (2.22) and (2.23).

The situation is slightly different for the ADM solutions (2.25) and the
real Ashtekar-like solutions (2.30).  Notice that the derivative
terms in (2.14) form a Laplace-Beltrami operator for the superspace metric
$$ g_{ij} = \pmatrix {{{1}\over {M^2}}& -{{1}\over {MN}}& -{{1}\over
{ML}}\cr -{{1}\over {MN}}& {{1}\over {N^2}}& -{{1}\over {NL}}\cr
-{{1}\over {ML}}& -{{1}\over {NL}}& {{1}\over {L^2}}\cr}, \eqno (3.4)$$
so we can define a conserved superspace current
$$j_0^k = -i\pmatrix {MN\Psi^*\partial_N \Psi + ML\Psi^* \partial_L \Psi -
MN\Psi \partial_N \Psi^* - ML\Psi \partial_L \Psi^*\cr MN\Psi^*
\partial_M \Psi + NL\Psi^* \partial_L \Psi - MN\Psi \partial_M \Psi^* -
NL\Psi \partial_L \Psi^*\cr ML\Psi^* \partial_M \Psi + NL \Psi^*
\partial_N \Psi - ML\Psi \partial_M \Psi^* - NL\Psi \partial_N \Psi^*\cr}
\eqno (3.5)$$
valid for any solution to (2.14).  Unfortunately $\Psi = We^{-S}$ for $W =$
const. and $S$ given by (2.25) is
{\it not\/} a solution to (2.14).  For $b = c = d = 0$ Moncrief and
Ryan\cite{MRY} showed that the factor ordering for
the derivative terms of (2.14) that
allows $W =$ const. is
$${{1}\over {L}} \partial_M MNL \partial_N + {{1}\over {N}}\partial_L
MNL \partial_M +  {{1}\over {M}}\partial_N MNL
\partial_L, \eqno (3.6)$$
and a conserved superspace current for this
equation is
$$j^k = (MNL)j^k_0, \eqno (3.7)$$
where $j^k_0$ is defined above.  Unfortunately both these currents are zero
for $\Psi$ given by (2.25) with $b = c = d = 0$, since $\Psi$ is real.  It
might appear that the solutions (2.30) would have nonzero currents since they
are complex, but the equation of motion (2.20) has pure imaginary terms, and
we have been unable to find a conserved current that is compatible with
it, so we cannot say that such a current would be nonzero.  The only
interesting ``probability'' measure is then the Hartle--Hawking
$|\Psi|^2$ which
we have given in Figs. 1-3.

We might worry about the ``normalization'' of the new functions given
by (2.25) with $b$, $c$, $d \neq 0$ since
they do not fall off rapidly for large $\beta_{\pm}$ (we are thinking of
``normalization'' in the sense of an ADM equation solution with $\alpha$
taken as an internal time, which means that $\int d\beta_{+} d\beta_{-}
|\Psi|^2$ over the $\beta_{+}\beta_{-}$-plane should be finite).
However, the ``arms'' of these solutions that remain finite as we move
out the Taub channels begin near $\beta_{+} = \beta_{-} = 0$ and become
very narrow rapidly, and the fact that
$$\int^{\infty}_0 d\beta_{+} \int^{\infty}_{-\infty} d\beta_{-} \exp
\{-12 e^{2\beta_{+}}\beta_{-}^2 \} = {{1}\over {2}} \sqrt{\pi/3} \eqno
(3.8)$$
implies that $\int d\beta_{+} d\beta_{-} |\Psi|^2$ will remain finite.

\section{Conclusions and Suggestions for Further Research}

In this paper we have studied the use of canonical transformations to
find particular solutions to quantum gravity, at least in the
minsuperspace sector.  What we showed was that any solution, $S$, to the
Einstein-Hamilton-Jacobi equation can be used to generate a canonical
transformation that leads to a solution of the form $\Psi = We^{-S}$.
The prefactor $W$ serves to
allow us to adjust the factor ordering of the Wheeler-DeWitt equation,
since any choice of $W$ is valid for some factor ordering.  For some
solutions $S$ the factor ordering that allows $W =$ const. is
relatively simple, as in (3.6), but for others, the only such factor
ordering we have found is somewhat clumsy (see Appendix B), but
simpler expressions for $W$ probably exist.

We also showed that there are, in principle, a large number of functions
$S$ that are solutions to modified Einstein-Hamilton-Jacobi equations
that represent other canonical transformations that do not annihilate the
curvature term in the Hamiltonian constraint.  Each of these canonical
transformations alows us to find exact solutions of the form $We^{-S}$.
The problem is that one would like to have an analytic solution for $S$,
and, in principle, one must solve a nonlinear functional differential
equation such as (1.5) in order to find $S$.  If one does not insist that
$S$ be a true solution to the Einstein-Hamilton-Jacobi equation, but
rather a
solution to the modified equation that would come from
(1.9), it may still be possible to
simplify the curvature term in the equation to the point where $S$ may be
found easily.

Seen from this viewpoint, the Ashtekar variables, based on a complex
canonical transformation, have the advantage that $S$ is known exactly
for the transformation, and one can construct solutions of the form
$We^{-S}$ with $W =$ const. easily.  Unfortunately there are two
drawbacks to this procedure.  One is that often the $S$ one calculates is
zero (for example, $S = 0$ is
a possible solution to Eqs. [2.22] and [2.23] with $\Lambda =
0$).  The other is that since the
transformation is complex, there is a need for a ``reality condition''
on the functions $\Psi$.  Because of this we investigated a real
transformation similar to the one that generates the Ashtekar variables
that requires no reality condition, but makes it necessary to solve a
more complicated equation for $S$.

In order to give concrete examples of the idea of finding solutions
given by
generators $S$ obtained
by solving the equivalent of the Einstein-Hamilton-Jacobi equation we
invesigated Bianchi-type minisuperspaces, and by writing the equations in
terms of variables of the Ashtekar type, we were able to find several
solutions for $S$, some of them new.  Unfortunately, the only
explicit factor ordering which allowed $W =$ const. we were
able to find is not very appealing.

Of course, minisuperspace solutions, while they may be of interest as
``wave functions of the universe'', are perhaps better thought of as
clues to finding solutions to full quantum gravity.  The Chern-Simons
solution in quantum gravity was found in just this way.  In the future we
might hope to find solutions to full quantum gravity that are suggested
by the minisuperspace exact solutions given above.  This would require
solutions $S$ to the functional Einstein-Hamilton-Jacobi equation or its
analogues where the curvature term is not annihilated.

This idea is perhaps the strongest suggestion for further research.  It
has proved fruitful to look for solutions to the Einstein-Hamilton-Jacobi
equation and its analogues, since it has been possible to find a number
of exact solutions to a problem such as the Mixmaster model which has a
reasonably complicated structure, and one could easily have assumed that
it would be impossible to find any analytic particular solutions to the
problem.  That it is fairly easy to find such solutions in this
minisuperspace case leads one to believe that it would not be impossible
to find such solutions in full quantum gravity.  The existence of the
Chern-Simons solution seems to point in this direction.
\vskip 1 true cm

\noindent {\bf ACKNOWLEDGMENTS}
\vskip 10 pt

We wish to thank C. Misner, T. Jacobson, B-L. Hu, and A. Vilenkin for
helpful discussions.  We also thank A. Ramirez for valuable help with
the figures.  One of us (M.R.) received support from the National
University of Mexico (DGAPA-UNAM) and the other (F.B.) from the
Spanish Research Council (CSIC), and we are both grateful for support
provided by NSF grant PHY93-96246 and the Eberly Research Fund of the
Pennsylvania State University.

\appendix
\section{Equations for $S$ and $W$}

The equations obtained by writing the wave function as $\Psi=W e^{-S}$ and
substituting it in (2.14, 2.16, 2.18, 2.20) are
$$\begin{array}{l}
-2\left[MN\partial_{M}\partial_{N}W+LM\partial_{L}\partial_{M}W+LN\partial_{L}
\partial_{N}W-\right.\\
-MN(\partial_{M}\partial_{N}S)W-LM(\partial_{L}\partial_{M}S)W-LN(\partial_{L}
\partial_{N}S)W-\\
-MN\partial_{M}S\partial_{N}W-ML\partial_{M}S\partial_{L}W-
     LN\partial_{L}S\partial_{N}W-\\
     -MN\partial_{M}W\partial_{N}S-ML\partial_{M}W\partial_{L}S-
     LN\partial_{L}W\partial_{N}S+\\
     \left.MN(\partial_{M}S\partial_{N}S)W+ML(\partial_{M}S\partial_{L}S)W+
     LN(\partial_{L}S\partial_{N}S)W\right]+\\

\left[\zeta(M^{2}+N^{2}+L^{2})-\frac{1}{2}\zeta\left(\frac{M^{2}L^{2}}{N^{2}}+
     \frac{N^{2}L^{2}}{M^{2}}+\frac{M^{2}N^{2}}{L^{2}}\right)-2\Lambda
     MNL\right]W=0
\end{array}
\eqno(A.1)$$
$$\begin{array}{l}
-(\alpha\beta+\gamma)\left[\partial_{\alpha}\partial_{\beta}W-(\partial_{\alpha}
    \partial_{\beta}S)W-\partial_{\alpha}S\partial_{\beta}W-\partial_{\alpha}W
    \partial_{\beta}S+(\partial_{\alpha}S\partial_{\beta}S)W\right]-\\

-(\alpha\gamma+\beta)\left[\partial_{\alpha}\partial_{\gamma}W-(\partial_{\alpha}

\partial_{\gamma}S)W-\partial_{\alpha}S\partial_{\gamma}W-\partial_{\alpha}W
    \partial_{\gamma}S+(\partial_{\alpha}S\partial_{\gamma}S)W\right]-\\

-(\beta\gamma+\alpha)\left[\partial_{\gamma}\partial_{\beta}W-(\partial_{\gamma}
    \partial_{\beta}S)W-\partial_{\gamma}S\partial_{\beta}W-\partial_{\gamma}W
    \partial_{\beta}S+(\partial_{\gamma}S\partial_{\beta}S)W\right]+\\
   +i\Lambda\left[\partial_{\alpha}\partial_{\beta}\partial_{\gamma}W-
    (\partial_{\alpha}\partial_{\beta}\partial_{\gamma}S)W-
    (\partial_{\alpha}\partial_{\beta}S)\partial_{\gamma}W-
    (\partial_{\alpha}\partial_{\gamma}S)\partial_{\beta}W-\right.\\
    -(\partial_{\beta}\partial_{\gamma}S)\partial_{\alpha}W
    -(\partial_{\alpha}\partial_{\beta}W)\partial_{\gamma}S
    -(\partial_{\alpha}\partial_{\gamma}W)\partial_{\beta}S
    -(\partial_{\beta}\partial_{\gamma}S)\partial_{\alpha}W+\\
    +(\partial_{\alpha}\partial_{\beta}S)(\partial_{\gamma}S)W
    +(\partial_{\alpha}\partial_{\gamma}S)(\partial_{\beta}S)W
    +(\partial_{\beta}\partial_{\gamma}S)(\partial_{\alpha}S)W+\\
    \left.+\partial_{\alpha}S\partial_{\beta}S\partial_{\gamma}W+
     \partial_{\alpha}S\partial_{\gamma}S\partial_{\beta}W+
     \partial_{\beta}S\partial_{\gamma}S\partial_{\alpha}W-
     (\partial_{\alpha}S\partial_{\beta}S\partial_{\gamma}S)W\right]=0
\end{array}
\eqno(A.2)$$
$$\begin{array}{l}
    -AB\partial_{A}\partial_{B}W-AG\partial_{A}\partial_{G}W-BG\partial_{B}
    \partial_{G}W+\\
    +AB\partial_{A}W\partial_{B}S+AG\partial_{A}W\partial_{G}S+BG\partial_{B}W
    \partial_{G}S+\\
    +AB\partial_{A}S\partial_{B}W+AG\partial_{A}S\partial_{G}W+BG\partial_{B}S
    \partial_{G}W+\\
    +AB(\partial_{A}\partial_{B}S)W+AG(\partial_{A}\partial_{G}S)W+
    BG(\partial_{B}\partial_{G}S)W-\\
    -AB(\partial_{A}S\partial_{B}S)W-AG(\partial_{A}S\partial_{G}S)W-
    BG(\partial_{B}S\partial_{G}S)W+\\
    +i\left(AB\partial_{G}W+AG\partial_{B}W+BG\partial_{A}W\right)-\\
    -i\left(AB(\partial_{G}S)W+AG(\partial_{B}S)W+BG(\partial_{A}S)W\right)+
    \Lambda\;ABG\;W=0
\end{array}
\eqno(A.3)$$
$$\begin{array}{l}
    -AB\partial_{A}\partial_{B}W-AG\partial_{A}\partial_{G}W-BG\partial_{B}
    \partial_{G}W+\\
    +AB\partial_{A}W\partial_{B}S+AG\partial_{A}W\partial_{G}S+BG\partial_{B}W
    \partial_{G}S+\\
    +AB\partial_{A}S\partial_{B}W+AG\partial_{A}S\partial_{G}W+BG\partial_{B}S
    \partial_{G}W+\\
    +AB(\partial_{A}\partial_{B}S)W+AG(\partial_{A}\partial_{G}S)W+
    BG(\partial_{B}\partial_{G}S)W-\\
    -AB(\partial_{A}S\partial_{B}S)W-AG(\partial_{A}S\partial_{G}S)W
    -BG(\partial_{B}S\partial_{G}S)W+\\
    +i\left(AB\partial_{G}W+AG\partial_{B}W+BG\partial_{A}W\right)-\\
    -i\left(AB(\partial_{G}S)W+AG(\partial_{B}S)W+BG(\partial_{A}S)W\right)-
    \Lambda\;ABG\;W+\\
    +(A^{2}+B^{2}+G^{2})W-\frac{1}{2}\left(\frac{A^{2}B^{2}}{G^{2}}+
    \frac{A^{2}G^{2}}{B^{2}}+\frac{B^{2}G^{2}}{A^{2}}\right)W=0\nonumber
\end{array}
\eqno(A.4)$$
The equations for $W$ are then
$$\begin{array}{l}
MN\partial_{M}\partial_{N}W+LM\partial_{L}\partial_{M}W+LN\partial_{L}
     \partial_{N}W-\\

-MN(\partial_{M}\partial_{N}S)W-LM(\partial_{L}\partial_{M}S)W-LN(\partial_{L}
    \partial_{N}S)W-\\
    -MN\partial_{M}S\partial_{N}W-ML\partial_{M}S\partial_{L}W-
     LN\partial_{L}S\partial_{N}W-\\
    -MN\partial_{M}W\partial_{N}S-ML\partial_{M}W\partial_{L}S-
     LN\partial_{L}W\partial_{N}S=0
\end{array}
\eqno(A.5)$$
$$\begin{array}{l}
-(\alpha\beta+\gamma)\left[\partial_{\alpha}\partial_{\beta}W-(\partial_{\alpha}
    \partial_{\beta}S)W-\partial_{\alpha}S\partial_{\beta}W-\partial_{\alpha}W
    \partial_{\beta}S\right]-\\
-(\alpha\gamma+\beta)\left[\partial_{\alpha}\partial_{\gamma}W-(\partial_{\alpha}

\partial_{\gamma}S)W-\partial_{\alpha}S\partial_{\gamma}W-\partial_{\alpha}W
    \partial_{\gamma}S\right]-\\
-(\beta\gamma+\alpha)\left[\partial_{\gamma}\partial_{\beta}W-(\partial_{\gamma}
    \partial_{\beta}S)W-\partial_{\gamma}S\partial_{\beta}W-\partial_{\gamma}W
    \partial_{\beta}S\right]+\\
    +i\Lambda\left[\partial_{\alpha}\partial_{\beta}\partial_{\gamma}W-
    (\partial_{\alpha}\partial_{\beta}\partial_{\gamma}S)W-
    (\partial_{\alpha}\partial_{\beta}S)\partial_{\gamma}W-
    (\partial_{\alpha}\partial_{\gamma}S)\partial_{\beta}W-\right.\\
-(\partial_{\beta}\partial_{\gamma}S)\partial_{\alpha}W
    -(\partial_{\alpha}\partial_{\beta}W)\partial_{\gamma}S
    -(\partial_{\alpha}\partial_{\gamma}W)\partial_{\beta}S
    -(\partial_{\beta}\partial_{\gamma}S)\partial_{\alpha}W+\\
+(\partial_{\alpha}\partial_{\beta}S)(\partial_{\gamma}S)W
    +(\partial_{\alpha}\partial_{\gamma}S)(\partial_{\beta}S)W
    +(\partial_{\beta}\partial_{\gamma}S)(\partial_{\alpha}S)W+\\
\left.+\partial_{\alpha}S\partial_{\beta}S\partial_{\gamma}W+
     \partial_{\alpha}S\partial_{\gamma}S\partial_{\beta}W+
     \partial_{\beta}S\partial_{\gamma}S\partial_{\alpha}W\right]=0
\end{array}
\eqno(A.6)$$
$$\begin{array}{l}
   -AB\partial_{A}\partial_{B}W-AG\partial_{A}\partial_{G}W-BG\partial_{B}
    \partial_{G}W+\\
    +AB\partial_{A}W\partial_{B}S+AG\partial_{A}W\partial_{G}S+BG\partial_{B}W
    \partial_{G}S+\\
+AB\partial_{A}S\partial_{B}W+AG\partial_{A}S\partial_{G}W+BG\partial_{B}S
    \partial_{G}W+\\
+AB(\partial_{A}\partial_{B}S)W+AG(\partial_{A}\partial_{G}S)W+
     BG(\partial_{B}\partial_{G}S)W\\
+i\left(AB\partial_{G}W+AG\partial_{B}W+BG\partial_{A}W\right)=0\nonumber
\end{array}
\eqno(A.7)$$
where in each case $S$ must be a solution for the corresponding Hamilton-Jacobi
equation. Notice that the equations for $W$ that come from (A.3) and (A.4)
appear to be
the same, but the functions $S$ that appear in them are different because
they are
solutions to different Hamilton-Jacobi equations.

\section{Factor ordering for the ADM solutions}

If we want $W =$ const., one possible factor ordering for the
derivative terms in (2.14) is
$$-2\left ( MN{{1}\over {\rho_1}} \partial_M \rho_1 \partial_N + ML
{{1}\over {\rho_2}} \partial_M \rho_2 \partial_L + NL {{1}\over
{\rho_3}} \partial_N \rho_3 \partial_L \right ), \eqno (B.1)$$
where $\rho_1$, $\rho_2$, $\rho_3$ are functions of $M, N, L$.
Inserting $\Psi = W_0 e^{-S}$ into (2.14) with the factor ordering
(B.1) and $W_0 =$ const. gives (for $S$ given by [2.25])
$$a\left ({{MN}\over {L}} + {{LN}\over {M}} + {{LM}\over {N}} \right
) - {{aM}\over {\rho_1}} {{\partial \rho_1}\over {\partial M}} \left (
-{{LM}\over {N}} + {{LN}\over {M}} + {{MN}\over {L}} \right ) - $$
$$ - {{aL}\over {\rho_2}} {{\partial \rho_2}\over {\partial L}} \left
( {{LM}\over {N}} - {{LN}\over {M}} + {{MN}\over {L}} \right ) -
{{aN}\over {\rho_3}} {{\partial \rho_3}\over {\partial N}} \left (
{{ML}\over {N}} + {{NL}\over {M}} - {{MN}\over {L}} \right ) -$$
$$ - dN {{M}\over {\rho_1}}{{\partial \rho_1}\over {\partial M}}
- cM{{L}\over {\rho_2}} {{\partial \rho_2}\over {\partial L}} - bL
{{N}\over {\rho_3}} {{\partial \rho_3}\over {\partial N}} = 0. \eqno
(B.2)$$
This is one equation for the three unknowns $\rho_1$, $\rho_2$,
$\rho_3$, so a solution is always possible.  If, as an example, we
take $\rho_1 = NL$, $\rho_2 = MN$, we find that $\rho_3$ becomes
$$ \rho_3 = MNL \exp \left \{ {{2\left ({{MN}\over {L}} + bL -
{{M}\over {L}} \right )}\over {\sqrt{M^2 + (1 - b^2) L^2}}} \tan^{-1}
\left ( {{-bL/2a + MNL}\over {\sqrt{M^2 + (1 - b^2) L^2}}} \right )
\right \}.
\eqno (B.3)$$
This solution is not elegant, and there are probably more symmetric
solutions that would be simpler.

\vfill\eject
\renewcommand{\thepage}{}
\centerline{FIGURE CAPTIONS}
\vskip 1 true cm

\noindent Figure 1. The square of the wave function $\Psi = e^{-S}$
with $S$ given by (2.25) with $a = 1/2$, $b = c = d = e = 0$.  The solid
line is an equipotential of the Mixmaster potential.  This solution
is peaked around the Robertson-Walker universe which is represented
by $\beta_{+} = \beta_{-} = 0$.
\vskip 0.5 true cm

\noindent Figure 2. $|\Psi|^2 = e^{-2S}$ for $S$ given by (2.25) with
$a = 1/2$, $b = c = d = -1$, $e = 0$.  The solid line is an
equipotential of the Mixmaster potential.  While the wave function is
still peaked around the Robertson-Walker universe, it has ``arms''
that single out the Taub models represented by the ``channels'' in
the potential.
\vskip 0.5 true cm

\noindent Figure 3. $|\Psi|^2 = e^{-2S}$ for $S$ given by (2.25) with
$a = 1/2$, $b = d = 1$, $c = -1$, $e = 0$.  The solid line is again an
equipotential of the Mixmaster potential.  This solution is no longer
peaked about the Robertson-Walker model, and lies almost entirely
over the $\beta_{+} > 0$ portion of the Taub model represented by
$\beta_{-} = 0$.
\vfill\eject

\renewcommand{\thepage}{}
\centerline{\psfig{figure=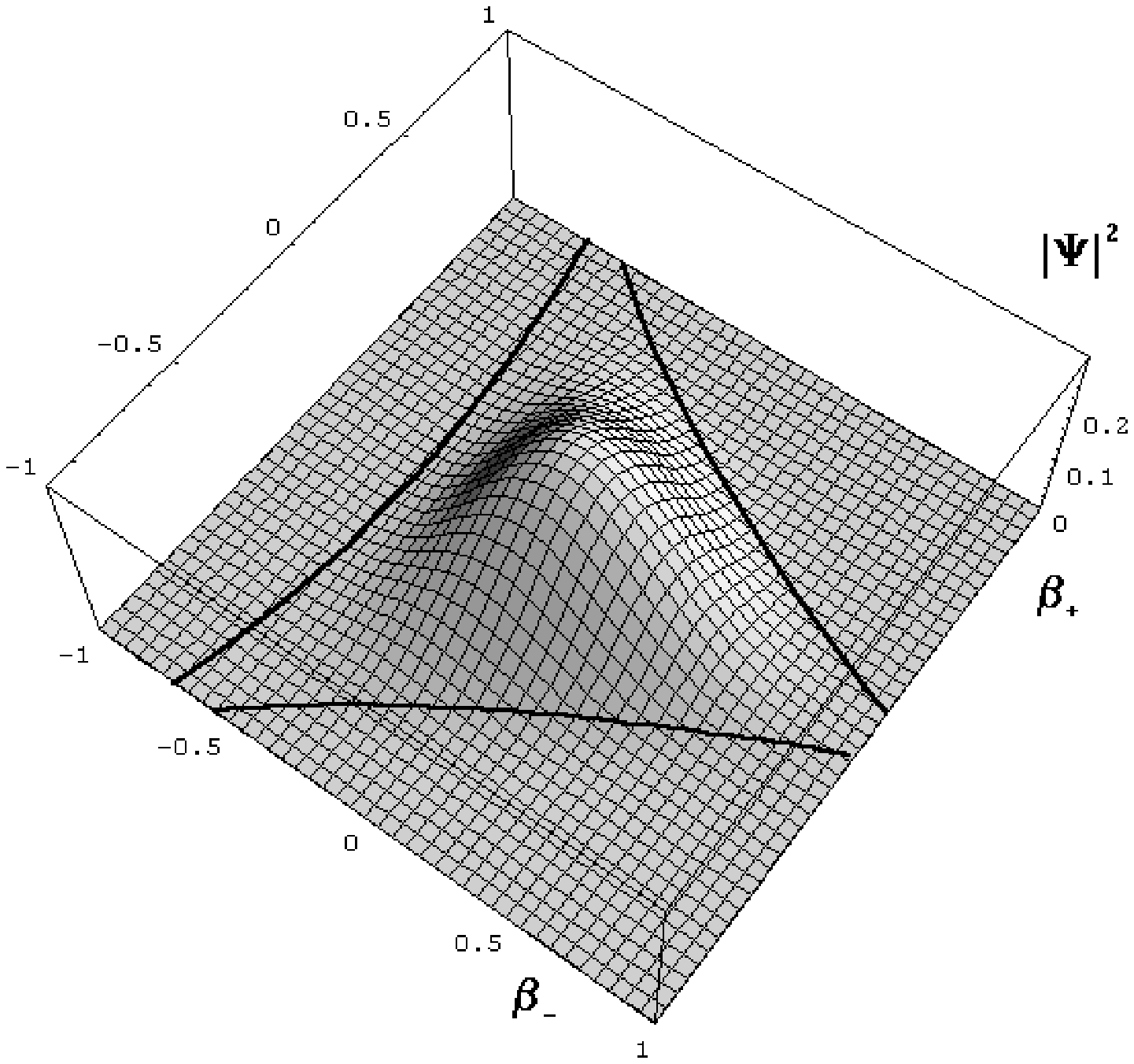,height=8in,bbllx=306bp,bblly=396bp,bburx=486bp,bbury=546bp,clip=}}
\centerline{Figure 1.}
\par
\vfill \eject
\renewcommand{\thepage}{}
\centerline{\psfig{figure=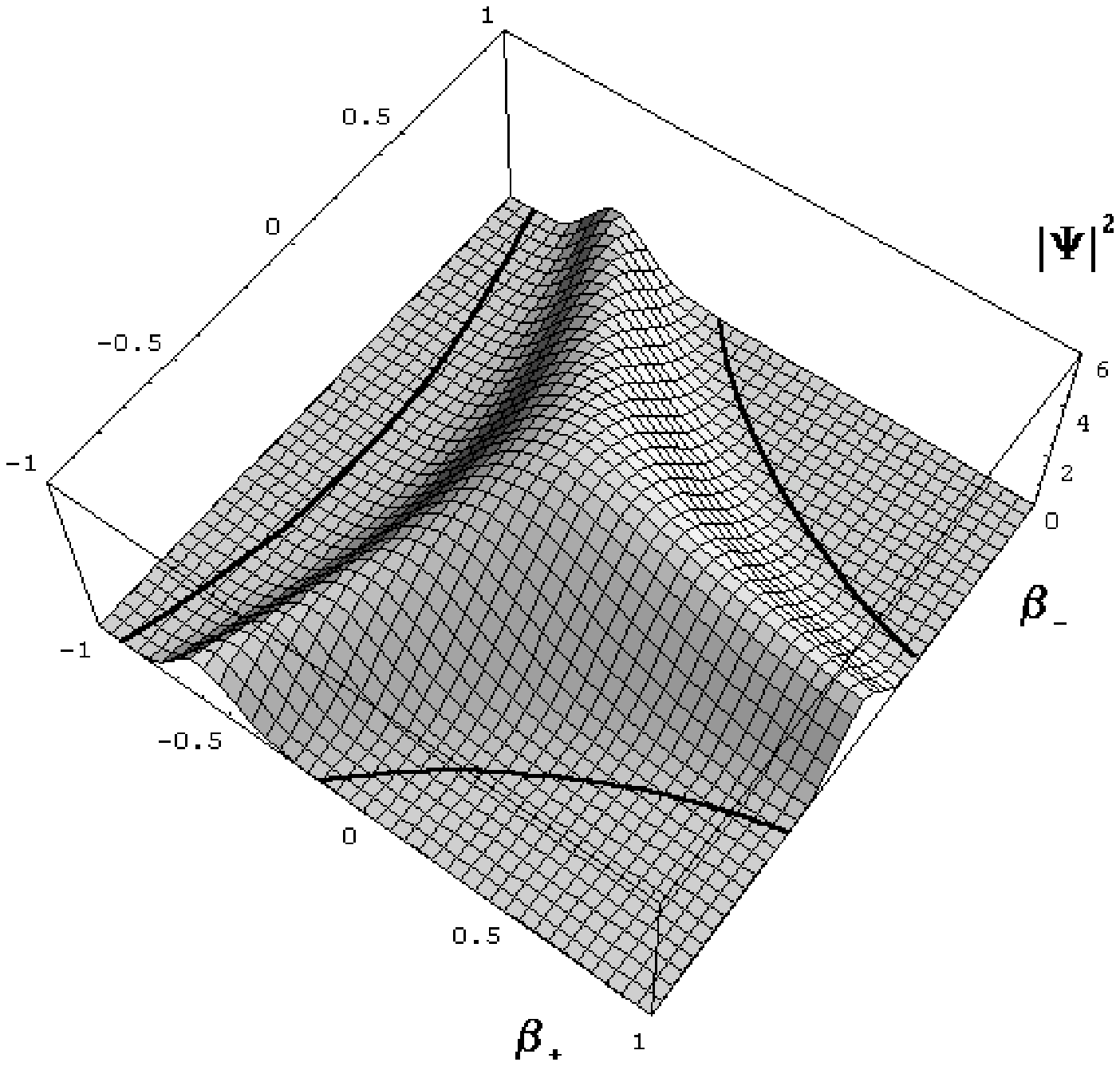,height=8in,bbllx=306bp,bblly=396bp,bburx=486bp,bbury=546bp,clip=}}
\centerline{Figure 2.}
\par
\vfill\eject
\renewcommand{\thepage}{}
\centerline{\psfig{figure=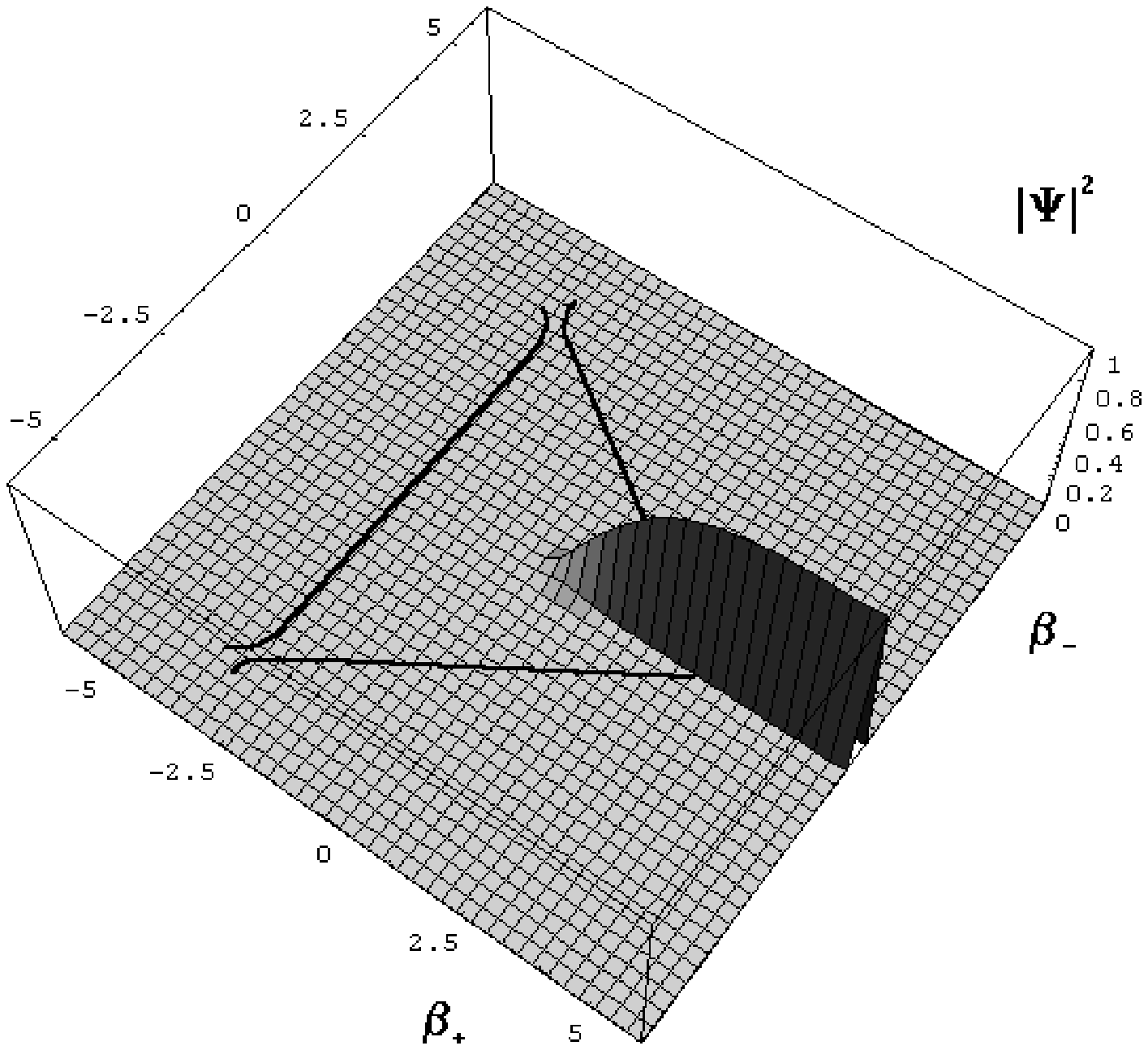,height=8in,bbllx=306bp,bblly=396bp,bburx=486bp,bbury=546bp,clip=}}
\centerline{Figure 3.}
\par
\vfill\eject
\end{document}